# Android Assistant EyeMate for Blind and Blind Tracker

Md. Siddiqur Rahman Tanveer, M.M.A. Hashem and Md. Kowsar Hossain
Department of Computer Science and Engineering
Khulna University of Engineering & Technology (KUET)
Khulna 9203, Bangladesh
sr.tanvir29@gmail.com, mma.hashem@outlook.com, auvikuet@yahoo.com

*Abstract*—At present many blind assistive systems have been implemented but there is no such kind of good system to navigate a blind person and also to track the movement of a blind person and rescue him/her if he/she is lost. In this paper, we have presented a blind assistive and tracking embedded system. In this system the blind person is navigated through a spectacle interfaced with an android application. The blind person is guided through Bengali/English voice commands generated by the application according to the obstacle position. Using voice command a blind person can establish voice call to a predefined number without touching the phone just by pressing the headset button. The blind assistive application gets the latitude and longitude using GPS and then sends them to a server. The movement of the blind person is tracked through another android application that points out the current position in Google map. We took distances from several surfaces like concrete and tiles floor in our experiment where the error rate is 5%.

## I. INTRODUCTION

Blind mobility is one of the main brainstorming challenges that scientists are still facing around different parts of the world and still researching to implement suitable blind assistive devices. In recent years blind mobility has become an important issue since a large number of people are visually impaired and partially sighted. According to the World Health Organization, approximately 0.4% of the population is blind in industrialized countries while the percentage is rising to 1% in developing countries [1, 2]. As of 2012 there were 285 million people who were visually impaired of which 246 million had low vision and 39 million were blind [3].

Navigating a blind person is a great challenge as blind person has to rely on other. The simplest and most widely used travelling aid used by all blinds is the white cane. It has provided those people with a better way to reach destination and detect obstacles on ground, but it cannot give them a high guarantee to protect themselves from all level of obstacles. Sometimes it happens that blind people are lost and their guardians are in tension about them. There has been many efforts but even now, it is not easy for the blind people to move independently from one place to another [4]. To solve this great problem it has been studied by many researches about support instruments for eye-sight [5].

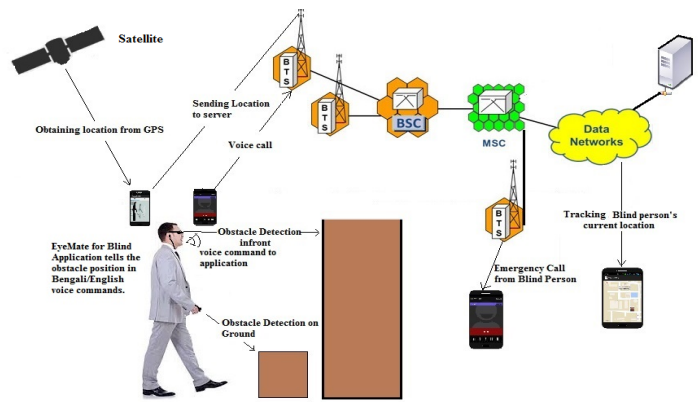

Fig. 1. Proposed solution to navigate and track Blind person.

According to the problems stated earlier we have proposed a system in this paper where a blind person can move without the help of other and can make emergency call to a predefined number and we can find out him/her easily if he/she is lost. The whole system is shown in Fig. 1. In our system we have found better performance in several surfaces during experimental phase where we found 5% error rate. The later part of this paper shows background information, high level system design, technical implementation, user experiences, and experimental results.

## II. BACKGROUND INFORMATION

### A. Problem Definition

Of all sensations perceived through our senses, those received through sight have the greatest influence on perception. Sight combined with the other senses, basically hearing, allow us to have a world global perception and to perform actions according to it. For the blind, the lack of sight is a major barrier in daily living: information access, mobility, way finding, interaction with the environment and with other people, among others, are challenging issues. In fact, school and working-age blind have very high analphabet and unemployment rates. For example, in the US, the blind unemployment rate is around 75% while only 10% of the blind children receive instruction in Braille [6]. Despite this effort, a true is that most schools and employers cannot accommodate blind people. In consequence, the person who is blind and his/her family faces important socioeconomic constraints. As blind person has to depend on others for their day to day life so often in our society, that people consider them as a burden. And often it happens that blind people fall down and get hurt while walking in the road so they need any device that can inform them about the position of the obstacle situated ahead. In this sense white cane is the most used blind assist tool for the blind person. Also it is a common scenario that blind people are lost and as a result their guardians and relatives are in tension and worst is that many of them are never found.

*B. Existing Approaches*

In modern times, there are various types of assistive technologies invented which are currently available to assist visually impaired people [3]. The main concern that many researchers are trying is to navigate a blind or partially sighted person quickly and safely in an unfamiliar environment. And from this point of view various projects have been introduced in recent years after long tiring researches that include different technologies like GPS, RFID, Ultrasonic, Laser and GSM [7-9]. Normally researchers have oriented wearable technology in various devices in such a way where these devices are actually worn on the body such as: assistive devices worn on fingers, wrist, forearm, and head. The authors in [9] proposed a blind assistive tool Guidecane where ultrasonic sonars are mounted in a semicircular fashion in guide wheels. In Guidecane wheels are connected with a cane and when any obstacle is detected, the wheels mounted ahead change the direction of the path left or right, and also the module contained a GPS module through which it guides the user to a destination through voice output telling the current position of the blind person. An RFID and smartphone based blind aid is proposed in [10].Where the blind person is guided through RFID reader and his routing path is maintained through the map designed in the application running in a Windows phone. The system uses Dijkstra's shortest path algorithm for path planning of the user and it gives direction to the user as voice commands from the application. The authors in [11] proposed an android based embedded system that uses two ultrasonic sensors mounted in tactile. In this system when any object is detected a feedback is given through the vibrators used in the cell phone and blind person is informed of his/her current location by this application. A system is proposed in [12] where authors showed travelling aids through several mobile applications. A GSM-GPS based system is proposed in [13], where the system allows blind people to enter notes and control device operation via Braille capacitive touch keypad instead of sending SMS by entering the number and text. An emergency button triggers an SMS from the GSM module that will send the current location of the user to a remote phone number asking for help.

Based on these circumstances, we have implemented a system which enables the blind person to walk independently in the road avoiding obstacles, guided by the voice commands according to the various obstacle positions by "EyeMate for Blind" android application. Also user can use his/her voice command to the application such as for establishing voice call. And the system facilitates guardians to monitor the movement of the blind person and rescue him/her in any situation using "BlindTracker", another android application introduced in this system.

III. HIGH LEVEL SYSTEM DESIGN

The proposed system has Hardware, Android, GPS and web based system capable of assisting the blind and visually impaired without the help of sighted person. To achieve the final objectives the following sub objectives have been accomplished:

*A. Object Detection*

UltraSonic range finders are used to detect the position of the obstacle and to measure the distance between the sensor and the object. Ultrasonic rangefinder sensor generates sound wave in an ultrasonic range and the sound is reflected from the object surface and during this time interval the number of pulses generated are counted and the actual distance is measured. The ultrasonic rangefinders used are the Maxbotix EZ0 sonar. We used three of these rangefinders, two of them are mounted on the spectacle and one is used for the hand-mounted tactile sensor. The sensors mounted in the spectacle detect the objects in the left and right side of the blind person and the sensor mounted in the finger detects the object situated on the ground. With a view to alerting the blind person about the obstacle position three vibrating motors (used in cell phone) are used, two of them in the spectacle and the other in the finger ring. And these are connected with an Arduino mega 2560 board.

*B. Bluetooth Module*

In this system a Bluetooth module HC-05 is used to send the data about the location of detected object to the android application. Since Bluetooth module is very cheap and easy to interface with the microcontroller and android phone so we used this module to communicate between the hardware module and "Eyemate for Blind" android application.

*C. EyeMate for Blind Android Application*

This android application is interfaced with the hardware module. This application receives the data sent by the microcontroller, If any object is detected then the application speaks in Bengali/English language to the blind person about the position of the object so that blind person can avoid collision. It also gets the latitude and longitude using the GPS available in android phone dynamically switching between the GPS provider and the Network provider. And every 5 minutes it sends the location to the server. Emergency Number is saved in the application. Blind person can establish voice call to the saved number and also control the application using his voice command in English.

*D. Global Positioning System (GPS)*

In our system we used the GPS to track the location of the blind person. Since every android phone is capable of using

GPS we do not need extra GPS module to interface in our system. Using the GPS latitude and longitude of the blind person can be obtained in two ways i) GPS provider ii) Network provider. We implemented both, because when the blind person walks under the open sky then GPS provider gives the location result but GPS provider will be disconnected by the tall buildings, trees and inside of the building and the underground. So for this reason GPS provider cannot be used all the time, and Network provider is to be used when GPS provider is unavailable.

### E. Remote Server

To store the traversed location of the blind person to remote database we implemented a remote server. EyeMate for Blind application sends the latitude, longitude and time to the server. And BlindTracker application retrieves this information from the server.

### F. Voice Command

Since blind person does not know about the android interface, we have developed our "EyeMate for Blind" application in such a way that the blind person can control the application using his/her voice command. Blind person needs to give voice command in English. We have implemented the system such that blind person does not need to touch the phone to give voice command - just he/she needs to press the headset button and can give the voice command easily.

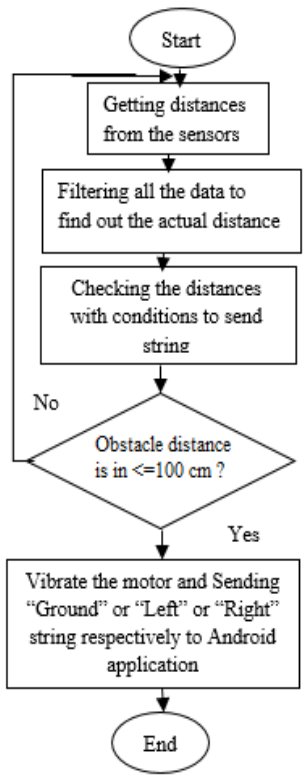

Fig. 2. Arduino program running algorithm.

### G. BlindTracker Android Application

This application is developed to track the current location of the blind person. Using this application we can first retrieve the latitude, longitude from server which is the current location of the blind person. Then find out the position in Google map. So, we can easily locate the blind person if he/she is lost and rescue him/her immediately.

### H. Google Map

With a view to tracking the position of the blind person immediately we need to point out the position in the map. So we have used the Google map to point out the current location of the blind person. In our project we used Google map using Google Map API v2.

### I. Physical Packaging

The implemented device is small and lightweight which makes it portable. Since we have implemented the hardware module in a spectacle so that the blind person can easily wear it and a sensor is mounted in finger ring so that blind person can detect the obstacle on ground.

## IV. TECHNICAL IMPLEMENTATION

### A. Object detection and Bluetooth Module Circuit

In our system the ultrasonic range finder is used. We can get data from range finder by three techniques. They are Serial, Analog and Pulse width modulation technique. The distance is measured according to the analog value derived from the sensor. It gives poor result and it deviates much from the actual one so pulse width modulation technique is applied here. According to this method per cm sensor generates 58 pulses and per inch 147 pulses. And its accuracy is good enough to measure accurate distance. We used the Arduino Mega 2560 board for implementing our project. The board contains ATmega 2560 microcontroller that operates these sensors and processes the filter operation of the achieved distance by the sonars and sends it to the android application via Bluetooth module. Three vibrating motors are connected with the board too.

### B. Software Design for Arduino Mega 2560 board

The flow chart shown in Fig. 2 represents the algorithm for the ATmega2560 microcontroller integrated in Arduino Mega 2560 board which processes the data and sends it to the android application via HC-05 Bluetooth module. Since we used the ultrasonic sound to measure the distance it is affected by several types of noise and the obtained distance is deviated from the actual distance so it is an important issue to find the best result. For this reason the median filter technique is used here. According to the technique 9 distances from same surface are taken using 10ms delay for sampling and then they are stored in array. To avoid unnecessary data while measuring distance if the measured value is greater than 15 and smaller than 645 then it is taken as good data. After getting 9 data the array is sorted and then using a median filter the actual distance is calculated. If any object is detected in the ground by the sensor situated in the finger ring which distance is smaller than

60 cm then the motor begins to vibrate situated alongside with finger ring and "Ground" string is sent to the EyeMate for Blind application via Bluetooth. Similarly if any object is detected in left or right by the sensors situated in the spectacle the motors situated there vibrate and then "Left" and "Right" strings are sent respectively to the application.

*C. Speaking the Obstacle Position*

EyeMate for Blind application is the main blind assistive application in this system. Through Bluetooth the application is interfaced with the hardware module. When it receives the string "Ground" and respectively "Left" and "Right" from Arduino board it tells in predefined Bengali or English voice commands to the blind person about the location of the obstacle. As a result the blind person can avoid that path. The flowchart of the application is shown in Fig. 3. The blind person can make voice call to a predefined emergency phone number saved in the application just by pressing the headset button and giving a simple voice command through English. Also he/she can control the application too. Every 5 minutes interval the application gets the latitude and longitude dynamically using the GPS provider or Network provider and sends the location information to the server with the current time.

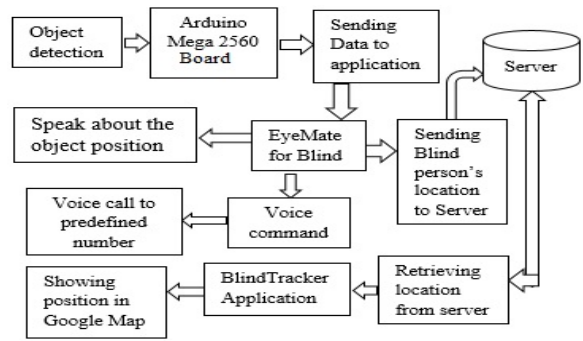

Fig. 4. System Block Diagram.

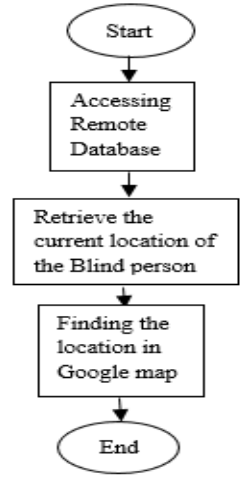

Fig. 5. BlindTracker application flowchart.

*D. Saving Latitude, Longitude and Time in Remote Server*

Since one portion of the system is to find out the location of a blind person through the BlindTracker application then we have implemented a remote server that contains a database where the latitude, longitude and time sent by the EyeMate for Blind application are saved. Server and android application interaction which is data parsing JSON (JavaScript Object Notation) is used here. The server and application interaction is shown in Fig. 4.

*E. Tracking the Location of Blind Person Using BlindTracker*

BlindTracker application gets the current inserted location by the EyeMate for Blind application from our server and then points out the location in the Google map. The flowchart of the Blind Tracker application is shown in Fig. 5.

## V. USER EXPERIENCES

*A. Alerting the Blind Person about the Obstacle Position*

We have implemented the device such a way that the sensors are mounted in a spectacle and in a finger ring. As a result the sensors mounted in the spectacle detect the obstacle position in front of the upper left or right side in the walking path and the finger ring mounted sensor detects the obstacle in the ground. The Fig. 6 shows the spectacle and ring.

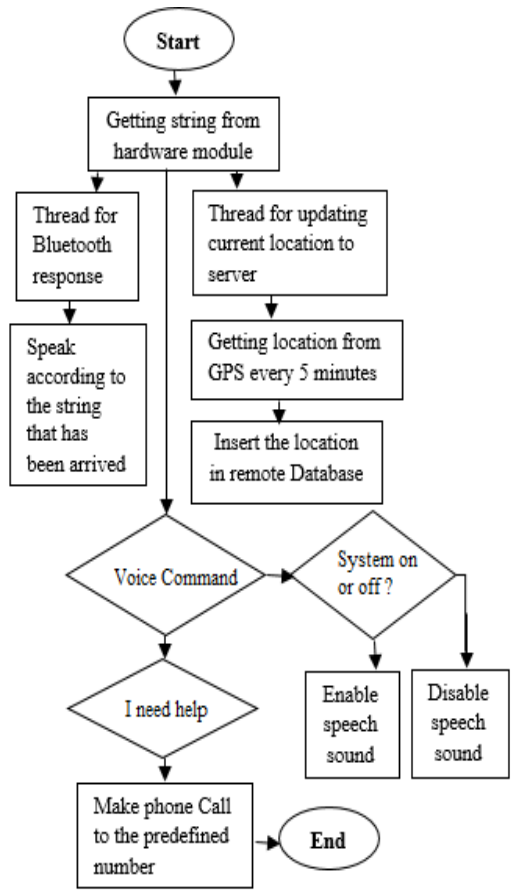

Fig. 3. EyeMate for Blind application algorithm.

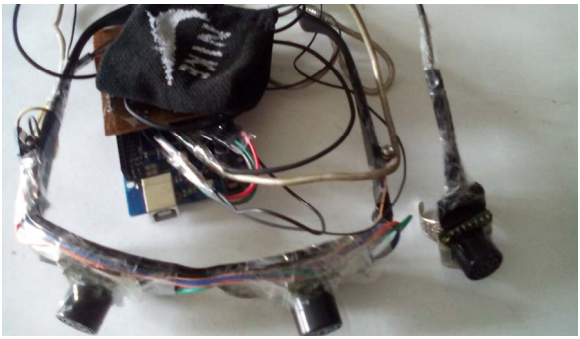

Fig. 6. Hardware module.

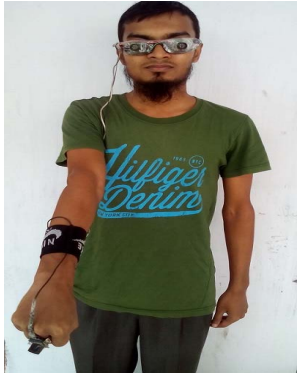

Fig. 7. User with Spectacle.

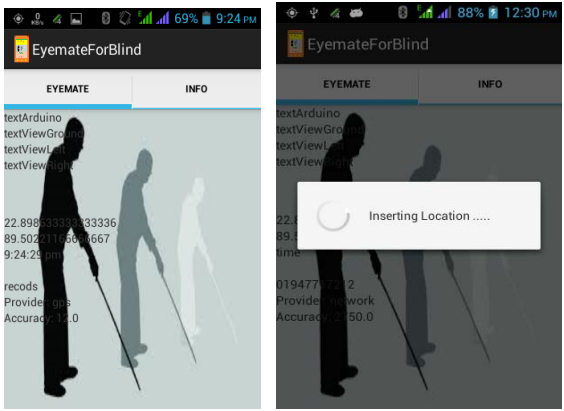

Fig. 8. Inserting current location to server.

A user wearing the device in his body is shown in Fig. 7. The interaction between the hardware module and EyeMate for Blind application is shown in Fig. 8 where the application tells the obstacle position to the blind person and in every 5 minutes the current is location measured by the GPS. Then latitude, longitude and time are sent to the server. As a result the traversed path of the blind person is saved in the server. Now the guardians can easily find out the currnt location of the blind person. As a result the current location of the blind person can be trcked through our BlindTracker tracking application. So guardians need not to worry about the blind person as they can easily know it from the server.

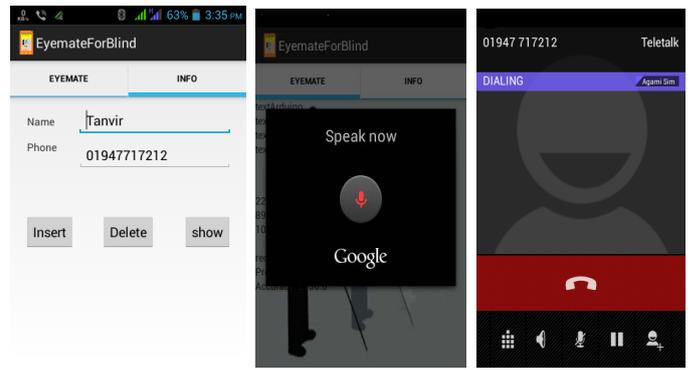

Fig. 9. Emergency Phone call establishment through voice command.

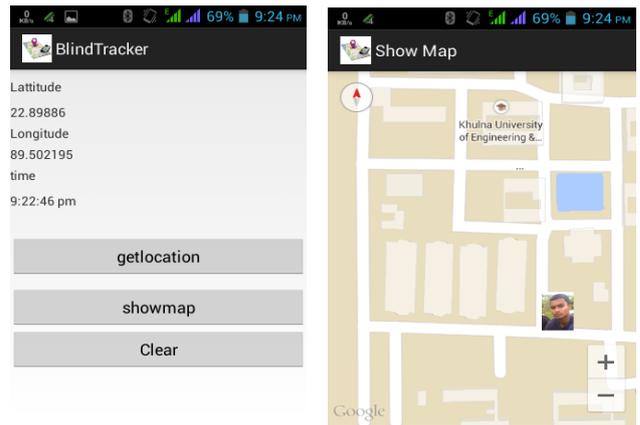

Fig. 10. Tracking Blind person's current position.

### B. Voice Command and Emergency Voice Call Establishment

As we have included the facility to save an emergency number in application so that the blind person can establish voice call to the predefined number by using his/her voice command. In Fig. 9 the voice command activities are shown. When the blind person wants to give a voice command, he/she needs not to touch the phone just pressing the button of the headset connected to the phone then a voice command prompt appears and then saying "I need help" this voice command will establish the voice call to the predefined number which is also represented in the figure. Also using the voice command a blind person can control the application if he/she does not want to hear the speech sound while sitting idle. And again he/she can activate the speech sound again.

### C. Finding the Current Location of the Blind Person

The Fig. 10 represents the tracking system to find out the current location of the blind person. Since it is our goal to reach to the blind person if he/she in danger, we have solved this challenge through this tracking appplication. At first we get the current location of the blind person just accessing the server just pressing the getlocation button in the BlindTracker application. Then the current latitude and longitude of the blind person are retrieved and then pressing the showmap button the currnt location of the blind person is shown in the map. As a result it has become very easy to help any blind person if the guardians get his/her emergency phone call or not.

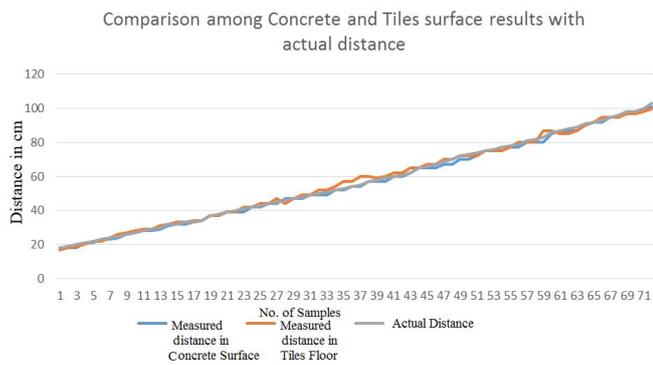

Fig. 11. Comparison among measured and actual distance in several surfaces

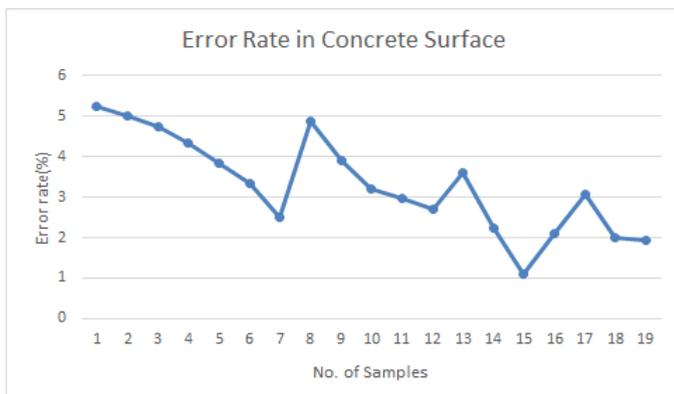

Fig. 12. Error Rate (%) in the measured distance

## VI. EXPERIMENTAL RESULTS

### A. Experimental Result Comparison Among Several Surfaces

We took several distances from various walking surfaces which were concrete, tiles floor. We performed our experiment in several places in both dry and wet weather. As a result we got some difference in the actual and the measured distance as sound wave generated by the sensors are affected much by humidity. We have presented a comparison graph between the measured distance and actual distance. Our comparison graph is shown in Fig. 11 where we have found that the distances measured in tiles floor give much better performance than the result found in the concrete surface. But in the wet surfaces the measured distances deviate much from the actual distance.

### B. Error Rate (%) in the Experimental Data

Our objective was also to implement the device so that the error rate should be minimum. During our experiment we took data from several noisy places like crowded road and shopping mall. As a result we got some error in our experiment. We have got 5% error in our experimental result which is shown in Fig. 12 as sound wave is affected much for noise and humidity.

## VII. CONCLUSION

In this paper, we have implemented a system for blind navigation. The final result of our approach is a blind navigation and tracking system with a flexible architecture that can be adopted in blind mobility. The system has been tested through several test cases. The blind assistive device with Eyemate for Blind android application is very useful for a blind person to move without the help of other and the user can seek emergency help through voice call. The tracking system involved here through BlindTracker application is very applicable to track the current location of the blind person.

REFERENCES

[1] Chaudhry M., Kamran M., Afzal S., "Speaking monuments — design and implementation of an RFID based blind friendly environment", 2nd ICEE International Conf. on 25-26 March, 2008.

[2] Velázquez R., "Wearable Assistive Devices for the Blind." Chapter 17 in A. Lay-Ekuakille & Mukhopadhyay (Eds.), Wearable and Autonomous Biomedical Devices and Systems for Smart ronment: Issues and Character ization, LNEE 75, Springer, 2010, pp 331-349.

[3] World Health Organization (2009) Visual impairment and blindness-Fact Sheet N°282. Available online at: http://www.who.int mediacentre/factsheets/fs282/en/

[4] C. Kang, H. Jo and B.Kim, "A Machine-to-Machine based Intelligent Walking Assistance System for Visually Impaired Person", The Journal of KICS, vol. 36, no. 3, pp. 195-304, 2011.

[5] N. Bourbakis, D. Dakopoulos, "A Comparative Survey on Wearable Systems for Blinds' Navigation," *1st International IEEE-BAIS Symposium on Research on Assistive Technologies*, Dayton, OH, pp.3-12, 16 April, 2007.

[6] Blind World Magazine (2006) Breaking the chains of paternalism. Available online at: http://home.earthlink.net/~blindworld/NEWS/6-06-14-02.htm

[7] Benjamin J. M., Ali N. A., and Schepis A. F., "A Laser Cane for the Blind." Proceedings of the San Diego Biomedical Symposium, Vol. 12, pp. 53-57.

[8] Madad A. Shah, Sayed H. Abbas, ShahzadA. Malik, " Blind Navigation via a DGPS-based Hand-held Unit", Australian Journal of Basic and Applied Sciences, 4(6): 1449-1458, 2010.

[9] Johann B. and Iwan U. "The GuideCane — A Computerized Travel Aid for the Active Guidance of Blind Pedestrians", Proceedings of the IEEE International Conference on Robotics and Automation, Albuquerque, NM, Apr. 21-27, 1997, pp. 1283-1288.

[10] Sandra M., Nik A. M., Maxim M., Aaron S. "BlindAid: An Electronic Travel Aid for the Blind", The Robotics Institute, Carnegie Mellon University, Pittsburgh, Pennsylvania 15213.

[11] Sachin B., Rohan T., Harshranga P., Bhurchandi, K. M. "Substitute eyes for Blind using Android" India Educators' Conference (TIIEC), Texas Instruments, 2013, pages: 38-43,DOI: 10.1109/TIIEC.2013.14

[12] Piotr K., Piotr S., Piotr W. and Piotr W. "Mobile Applications Aiding the Visually Impaired in Travelling with Public Transport", Proceedings of the 2013 Federated Conference on Computer Science and Information Systems, pp. 825–828

[13] B.R.Prudhvi and Rishab B. "Silicon Eyes: GPS-GSM based Navigation Assistant for Visually Impaired using Capacitive Touch Braille Keypad and Smart SMS Facility" Computer and Information Technology (WCCIT), World Congress on 2013, Pages: 1-3, DOI: 10.1109/WCCIT